\documentclass[natbib]{emulateapj}
\usepackage{natbib}
\usepackage{graphicx}
\usepackage{bm,color}
\input epsf
\def\s2n{S^{\prime}/N}

\def\pdv{p\,dV}
\def\gsim{\;\rlap{\lower 2.5pt
\hbox{$\sim$}}\raise 1.5pt\hbox{$>$}\;}
\def\lsim{\;\rlap{\lower 2.5pt
\hbox{$\sim$}}\raise 1.5pt\hbox{$<$}\;}

\newcommand{\Eq}[1]{Equation (\ref{eq:#1})}
\newcommand{\eq}[1]{Eq.\ (\ref{eq:#1})}
\newcommand{\eqs}[2]{Eqs.\ (\ref{eq:#1}) and (\ref{eq:#2})}

\usepackage{amssymb,amsmath}
\usepackage{mathrsfs}

\def\bs{\boldsymbol}

\begin{document}
\title{Inaccuracy of Spatial Derivatives in Simulations of Supersonic Turbulence}

\author{Liubin Pan$^1$}
\author{Paolo Padoan$^{2,3}$}
\author{{\AA}ke Nordlund$^4$}
\affiliation{$^1$School of Physics and Astronomy, Sun Yat-sen University, 2 Daxue Road, Zhuhai, Guangdong, 519082, China; panlb5@mail.sysu.edu.cn}
\affiliation{$^2$Institut de Ci\`{e}ncies del Cosmos, Universitat de Barcelona, IEEC-UB, Mart\'{i} Franqu\`{e}s 1, E08028 Barcelona, Spain; 
ppadoan@icc.ub.edu}
\affiliation{$^3$ICREA, Pg. Llu\'{i}s Companys 23, 08010 Barcelona, Spain}
\affiliation{$^4$Centre for Star and Planet Formation, Niels Bohr Institute and Niels Bohr Institute, University of Copenhagen, {\O}ster Voldgade 5-7, DK-1350 Copenhagen K, Denmark; aake@nbi.ku.dk}

\begin{abstract}
We examine the accuracy of spatial derivatives computed from numerical simulations of supersonic turbulence. 
Two sets of simulations, carried out using a finite-volume code that evolves the hydrodynamic equations with an approximate 
Riemann solver and a finite-difference code that solves the Navier-Stokes equations, are tested against a number of 
criteria based on the continuity equation, including exact results at statistically steady state. We find that the 
spatial derivatives in the Navier-Stokes runs are accurate and satisfy all the criteria. In particular, they satisfy 
our exact results that the conditional mean velocity divergence, $\langle \nabla \cdot {\bs u}|s\rangle$, where $s$ 
is the logarithm of density, and the conditional mean of the advection of $s$, $\langle  {\bs u} \cdot \nabla s|s\rangle$, 
vanish at steady state for all density values, $s$. On the other hand, the Riemann solver simulations fail all the tests that 
require accurate evaluation of spatial derivatives, resulting in apparent violation of the continuity equation, even if the solver
enforces mass conservation. In particular, analysis of the Riemann simulations may lead to the incorrect conclusion that the 
$\pdv$ work tends to preferentially convert kinetic energy into thermal energy, inconsistent with the exact result that the 
energy exchange by $\pdv$ work is symmetric in barotropic supersonic turbulence at steady state. The inaccuracy of spatial
derivatives is a general problem in the post-processing of simulations of supersonic turbulence with Riemann solvers. 
Solutions from such simulations must be used with caution in post-processing studies concerning  the spatial gradients.
\end{abstract}

\maketitle

\section{Introduction}

Numerical simulations are a powerful tool for turbulence studies and have driven major advances in our knowledge of highly 
compressible turbulence in the interstellar medium. Through visualization and statistical analysis, numerical simulations have 
helped capture and quantify the complicated density and velocity structures in supersonic interstellar turbulence, offering physical 
insight for the interpretation of observational results. In particular, numerical experiments have played a crucial role in the theoretical 
modeling of the process of star formation in molecular clouds (e.g. Padoan et al. 2014).   

The limited spatial resolution is still a major challenge for the interpretation of numerical simulations of turbulence, because the 
effective Reynolds number of even the largest simulations is still orders of magnitude lower than that in interstellar clouds. Many 
astrophysical simulations do not even reach a broad enough dynamical range of scales to allow the development of an inertial range
of turbulence (let alone to describe the turbulent fragmentation process down to the sonic scale). In order to maximize the inertial range, 
the strategy of choice in supersonic turbulence is to use Riemann solvers to evolve 
the integral form of the Euler equations rather than the Navier-Stokes equations, 
based on the assumption that the inertial-range dynamics is insensitive to the exact form of the viscosity (following Kolmogorov's universality hypothesis for the 
inertial-range properties of turbulence, e.g., Frisch 1995). 
In such simulations, energy dissipation occurs only implicitly, without an explicit viscous term, and the simulated flow may acquire 
an effective Reynolds number significantly larger than in corresponding Navier-Stokes simulations.  For convenience we refer
to such simulations as `Euler simulations' in this paper, even though the solutions of course---especially in the supersonic
context---would not exist without the implicit viscosity present in the Riemann solver, and `Navier-Stokes solutions with 
implicit sub-grid viscosity' would be a more precise label. These simulations are also referred to as implicit large-eddy simulations 
(ILES) in the numerical literature.

Without an explicit physical viscosity in Euler simulations, velocity and density profiles in sharp discontinuities are completely 
determined by the numerical algorithm (e.g. choice of Riemann solvers) and the specific methods to achieve numerical stability 
(e.g. choice of slope limiters). The resulting profile shapes within the discontinuities are numerically stable and 
the resulting solutions strictly obey the conservation laws, but with fluxes of mass, momentum, and total energy that are 
modified on the cell scale, relative to fluxes computed directly from the conserved variables. The extent to which this 
numerical approach affects diagnostics that rely on small-scale
spatial derivatives of the flow quantities has not been carefully addressed so far. The goal of this work is to quantify the
consequence of such inaccuracies in simulations of supersonic turbulence without explicit viscosity, which may have a significant 
impact for a large number of numerical studies.  

Spatial derivatives are indeed often needed in the statistical analysis of turbulence simulation data. 
The velocity gradient tensor and its three components, namely the rate of strain tensor, 
the vorticity, and the divergence are of great theoretical interest and have been extensively investigated in 
turbulence studies. In compressible turbulence, the statistics of vorticity and divergence are useful measures 
to characterize the solenoidal and compressible parts of the velocity field (e.g., Porter et al.\ 2002, Pavlovski et al.\ 2006, Wang et al.\ 2011). 
Furthermore, there is a wide range of important physical quantities related to the velocity gradient. Examples include, 
but are not limited to, the enstrophy, the Taylor microscale (Porter et al.\ 1992a,b, Kritsuk et al.\ 2007), 
the helicity (Porter et al.\ 1992b, Kritsuk et al.\ 2007), the energy dissipation rate (e.g., Kritsuk et al.\ 2007, Pan \& Padoan 2009, 
Pan et al.\ 2009), the small-scale compressive ratio (Kida \& Orszag 1990, 1992, Porter et al.\ 1992a,b, Kritsuk et al.\ 2007), 
the pressure-dilatation interactions or the $\pdv$ work (Porter et al.\ 1998, Pan \& Scannapieco 2010, Aluie et al.\  2012). 
The velocity gradient is also needed to study the preferential alignment of the vorticity with the principle directions of the strain tensor 
(e.g., Porter et al.\ 1998, Sur et al.\ 2013). The analysis of all the above physical quantities relies on accurate computation of the spatial 
derivative of the velocity field, yet many of the works cited above are based on Euler simulations, where velocity gradients may 
suffer from numerical artifacts. 

Although it should be straightforward to realize that spatial derivatives from simulations without explicit viscosity may be inaccurate 
at discontinuities, to our knowledge, the issue has so far not been carefully examined. It may have been overlooked because of the 
overwhelming interest in understanding the inertial-range dynamics of supersonic turbulence, rather than in achieving accurate small-scale 
statistics, or because of the lack of an objective criterion to quantify numerical errors in the computed spatial derivatives.  
In this work, we examine spatial derivatives in simulations of supersonic turbulence using a number of criteria, including exact results 
we have previously derived from the continuity equation at steady state (Pan et al.\ 2018). We show that the spatial derivatives in the
Euler simulations are not reliable, especially in dense regions. In contrast, in simulations that evolve the Navier-Stokes equations, spatial 
derivatives are found to be accurate. 

In \S 2, we outline the criteria used to test the accuracy of spatial derivatives in numerical simulations. 
\S 3 examines spatial derivatives in two sets of simulations, which are carried out using two different 
numerical codes with and without an explicit viscous term. Implications of our results are discussed in \S 4, 
and the main conclusions of the study are summarized in \S 5.

\section{Criteria to test spatial derivatives}

We start by writing down the continuity equation, 
\begin{equation}
\frac {\partial \rho }{\partial t}  +   \nabla  \cdot  (\rho \bs{u}  ) = 0,
\label{eq:rho}
\end{equation}
and the momentum equation, 
\begin{equation}
\frac {\partial (\rho  \bs{u}) }{\partial t}  +   \nabla  \cdot  (\rho \bs{u}  \bs{u}) = - \nabla p +  \nabla \cdot \mathcal{\sigma}   +\rho {\bs a},
\label{eq:v}
\end{equation}
where $\sigma$ denotes the viscous stress tensor, and ${\bs a}$ is the acceleration that drives the turbulent flow. 
All other symbols carry their conventional meanings. In most simulations of supersonic turbulence the viscous term, 
$\nabla \cdot \mathcal{\sigma}$, in the momentum equation is neglected, in order to minimize the thickness of discontinuities 
and achieve the largest possible effective Reynolds number. However, such sharp discontinuities are numerically unstable.
The necessary corrections to the profile of the flow variables within the discontinuities to obtain the desired stability, such as 
flux or slope limiters or various forms of explicit numerical dissipation yield inaccurate spatial derivatives in the discontinuities.

We will evaluate the uncertainties in the spatial derivatives computed from simulation data using a number of criteria.  
A simple test is to check whether and how well the continuity equation is satisfied with density and velocity 
derivatives computed from the data. For convenience, we rewrite \eq{rho} as,    
\begin{equation}
\frac {\partial s}{\partial t}  + \bs{u}  \cdot \nabla s + \nabla \cdot \bs{u}=0,
\label{eq:s}
\end{equation}
where $s\equiv \ln (\rho/\bar{\rho})$ is the logarithm of the density, $\rho$, with $\bar{\rho}$ being the mean density.

In a perfect numerical simulation, one would expect that \eq{s} is exactly satisfied. 
It may seem trivial that, as long as mass conservation is enforced in the simulation, \Eq{rho} and, by inference, 
\Eq{s} should hold exactly. That is, however, not true, if the numerical schemes
involve flow corrections designed to stabilize shocks, as such corrections introduce errors in the spatial 
derivatives, even while enforcing mass conservation.  
As a consequence, the density and velocity fields 
from an Euler simulation may appear to not satisfy the continuity equation, \Eq{rho}, and the three terms in \eq{s}
may not add up to zero when evaluated with finite-difference derivatives of density and velocity, even if the code strictly enforces mass conservation. 
This apparent contradiction
has a simple explanation: a generic finite-volumes Euler code does not require those spatial derivatives 
to evolve the flow solution (the code does not solve the continuity equation in the differential form of \eq{s}, 
but solves an integral form of \eq{rho}), so the solution is evolved correctly and satisfies the conservation 
laws despite the incorrect spatial derivatives. 

To quantify this potential problem, we consider the sum of the three terms of the continuity equation, which we refer to as the 
numerical or artificial residual of the continuity equation and denote as $R= \partial_t s + \bs{u}  \cdot \nabla s +\nabla \cdot \bs{u}$. 
In the following section, we check whether $R$ is zero point-wise on 
the simulation grid, and, if not, find out where deviations typically occur. We also test statistically how well \eq{s} 
is satisfied in flow regions at different density. The test is carried out by computing the averages of the three terms in \eq{s}, conditioned on the local flow density, denoted as 
$\langle  \partial_t s|s\rangle$, $\langle \nabla \cdot \bs{u}| s \rangle $ and $\langle \bs{u}  \cdot \nabla s| s \rangle$, respectively. 
Here $\langle \cdot \cdot \cdot |s\rangle$ represents the average over the flow regions at a given value of $s$. 
Obviously, if \eq{s} held perfectly, we would have $\langle R|s\rangle =0$ or,
\begin{equation}
 \langle \partial_t s|s\rangle + \langle   \bs{u}  \cdot \nabla s|s \rangle + \langle \nabla \cdot \bs{u} |s\rangle =0,
\label{eq:scondition}
\end{equation}
at each density level. A test based on this equation can tell us at which density the spatial derivatives computed from a simulation
are more (or less) reliable.  

In addition to directly testing spatial derivatives against the continuity equation, we also 
consider several other criteria based on exact results derived from the continuity equation at 
statistically steady state. We outline such criteria in the following subsections.

\subsection{Time derivative of $s$}

At statistically steady state, the average, $\langle s \rangle$, of $s$ becomes time-independent, and it is expected that 
$\langle \partial_t s \rangle = \partial_t \langle s \rangle =0$. The assumption of steady state has stronger 
implications than $\langle \partial_t s \rangle =0$.  In fact, assuming that the probability distribution of $s$ is invariant with 
time at steady state, one can show that the conditional average of $\partial_t s$ on the 
density is zero at each density level. Our proof begins with defining a fine-grained PDF 
of $s$ as $g(\zeta;  {\bs x}, t) = \delta(\zeta - s({\bs x}, t))$,  where $\delta$ is the Dirac delta function and $\zeta$ 
the sampling variable. The time derivative $g$ is given by,
\begin{equation}
\frac{ \partial g (\zeta; {\bs x}, t)}{\partial t} = -\frac{\partial ( g \partial_t s)}{\partial \zeta},
\label{eq:sfine}
\end{equation} 
where we used the fact that $\partial_t s$ is independent of the sampling variable $\zeta$ for the term on the right hand 
side.
 
We then define the coarse-grained PDF as the average of the fine-grained PDF, i.e., $f(\zeta; t) 
\equiv \langle g(\zeta; {\bs x}, t) \rangle = V^{-1} \int_V \delta(\zeta - s({\bs x}, t) d^3x$, 
where the integral is over the volume, $V$, of the entire flow domain.  For any flow quantity $\phi ({\bs x}, t)$, it can be shown 
that $\langle  \phi ({\bs x}, t) g(\zeta; {\bs x}, t) \rangle = \langle  \phi ({\bs x}, t) \delta(\zeta - s({\bs x}, t)) \rangle = \langle  \phi | s = \zeta \rangle f(\zeta; t)$, 
where $\langle  \phi | s= \zeta \rangle$ is the average of $\phi$ over all the flow regions where $s( {\bs x}, t)$ equals the sampling variable (Pope 2000).
Averaging \eq{sfine} then gives,  
\begin{equation}
\frac {\partial f (\zeta; t)}{\partial t} = - \frac{\partial ( \langle  \partial_t s|s=\zeta \rangle f)}{\partial \zeta}, 
\label{eq:scoarse}
\end{equation}
and further assuming steady state, we have $\partial_\zeta ( \langle  \partial_t s|s=\zeta \rangle f) =0$. 
Therefore, $\langle  \partial_t s|s=\zeta \rangle f =C_1$, with $C_1$ the integration constant.  
Considering that  $\int_{-\infty}^{\infty} \langle  \partial_t s|s=\zeta \rangle f d\zeta= \langle \partial_t s \rangle$, which vanishes at steady 
state, the constant $C_1$ must be zero, and we obtain,  
\begin{equation}
\langle  \partial_t s|s=\zeta \rangle = 0, 
\label{eq:exact-relation0}
\end{equation}
meaning that the conditional mean of the time derivative of $s$ is zero at any density. 
The only assumption made in the derivation is that the coarse-grained  PDF is time independent, i.e., $\partial_t f (\zeta; t) =0$, 
which is expected once the flow reaches steady state. \Eq{exact-relation0} will be used to 
check whether the simulated flow has reached steady state.  

For simplicity of notations,  in the rest of the paper, we will drop the sampling variable, 
$\zeta$, in the conditional means, and write $\langle ...|s =\zeta\rangle $ simply as $\langle ...|s\rangle $. 
The density PDF will be written as $f(s; t)$ accordingly. 
    
\subsection{Exact results of Pan et al.\ (2018)}

Pan et al.\ (2018) derived two exact results from the continuity equation under the 
assumption of statistical stationarity and homogeneity. We briefly 
review the derivation here and refer the interested reader to Pan et al.\ (2018) for more details. 

At steady state, it follows from \eqs{scondition}{exact-relation0}  that,
\begin{equation}
\langle   \bs{u}  \cdot \nabla s|s \rangle + \langle \nabla \cdot \bs{u} |s\rangle=0.
\label{eq:coarse}
\end{equation} 
In Pan et al.\ (2018), the two terms were referred to as the conditional mean advection and conditional mean divergence, respectively. 
Using periodic boundary conditions (which is equivalent to the assumption of statistical homogeneity in Pan et al.\ 2018), 
the two terms are related to each other by\footnote{The relation is derived by 
averaging the identity $g\nabla \cdot \bs{u} = \nabla \cdot (g\bs{u} )- \bs{u} \cdot \nabla g  =  \nabla \cdot (g\bs{u} )+ \partial_{\zeta} ( g \bs{u} \cdot   \nabla s ) $ (see Pan et al.\ 2018).}, 
\begin{equation}
\langle \nabla \cdot \bs{u}| s \rangle f = \frac{\partial }{\partial s} \Big( \langle \bs{u} \cdot \nabla s |s  \rangle f \Big).
\label{eq:div-adv-relation}
\end{equation} 
Combining \eqs{coarse}{div-adv-relation} yields, 
\begin{equation}
\frac {\partial} {\partial s}  \Big(\langle \nabla \cdot \bs{u} |s\rangle f\Big) +\langle \nabla \cdot \bs{u}| s \rangle f  =0, 
\end{equation} 
which is solved by $\langle \nabla \cdot \bs{u}| s \rangle f (s) = C_2 \exp(-s)$ with $C_2$ being the integration constant. 
Because $\int_{-\infty}^{\infty} \langle \nabla \cdot \bs{u}| s \rangle f (s) ds = \langle \nabla \cdot \bs{u} \rangle =0$, the constant 
$C_2$ must vanish, and we have, 
\begin{equation}
\langle \nabla \cdot \bs{u}| s \rangle = 0,
\label{eq:exact-relation1}
\end{equation} 
meaning that the divergence in expanding and contracting regions at any given density exactly  cancels out. 

Combining \eqs{div-adv-relation}{exact-relation1} yields $\langle \bs{u} \cdot \nabla s |s  \rangle f =C_3$ where $C_3$ is another integration 
constant.   The integral of $\langle \bs{u} \cdot \nabla s |s  \rangle f$ over the $s-$space 
is equal to   $\langle \bs{u} \cdot \nabla s \rangle $. By averaging \eq{s} and using statistical homogeneity,  
it is straightforward to see that $\langle \bs{u} \cdot \nabla s\rangle =0$ at steady state. This requires $C_3=0$, leading to the second exact result of Pan et al.\ (2018), 
\begin{equation}
\langle \bs{u}  \cdot \nabla s| s \rangle = 0.
\label{eq:exact-relation2}
\end{equation} 

Pan et al.\ (2018) used the exact results, \eqs{exact-relation1}{exact-relation2}, to test potential numerical artifacts in Euler simulations 
of supersonic turbulence. In \S 3, we will further test the spatial derivatives in simulated turbulent flows using both Euler and Navier-Stokes simulations.  

We stress that the derivation in \S 2.1 for $\langle \partial_t s|s\rangle=0$ only assumed steady state, while here the derivation of 
\eqs{exact-relation1}{exact-relation2} requires both statistical stationarity and homogeneity (through the use of periodic boundary conditions). 
The results in \S 2.1 and 2.2 show that, if the continuity equation is perfectly satisfied, each of the three terms in \eq{scondition} is 
zero at steady state. On the other hand, if the continuity equation does not hold exactly, \eq{scondition} breaks down, 
and at least one of the three terms is nonzero. Each of the three terms will be examined in \S 3. 
 
\subsection{The $\pdv$ work}

Kinetic energy and thermal energy in a turbulent flow may be converted into each other through $\pdv$ work.
Unlike viscous dissipation, which is a one-way conversion of kinetic energy into heat, 
the energy exchange by $\pdv$ work is reversible, and its rate $p \nabla \cdot {\bs u}$ 
can be either positive or negative. An interesting question is whether the two-way energy exchange by $\pdv$ work  is symmetric in 
supersonic turbulence. In other words, does the energy conversion by $\pdv$ work have a 
preferred direction, causing a net energy flow from one form of energy to the other? To answer the question, one 
may compute the average $\pdv$ work, $\langle p \nabla \cdot {\bs u} \rangle$, over the flow domain 
and check if it is positive, negative or zero.  Clearly, due to its dependence on the 
divergence, an accurate evaluation of $\pdv$ work requires reliable spatial derivatives of the velocity field.  

Previous numerical studies found that the energy exchange by $\pdv$ work in isothermal, supersonic turbulence 
is asymmetric. At steady state, $\langle p \nabla \cdot {\bs u} \rangle$ was found to be negative, 
meaning that the $\pdv$ work preferentially converts kinetic energy to heat (Pan \& Scannapieco 2010, see also Kritsuk et al.\ 2013). 
However, this conclusion is questionable because the simulations used in those  
studies do not include explicit viscosity, 
and, as discussed earlier, the velocity divergence in such simulations may suffer from numerical artifacts. 

\subsubsection{Energy exchange by $\pdv$ work is symmetric}

In fact, under the assumption of statistical homogeneity, one can show that 
$\langle p \nabla \cdot {\bs u} \rangle$ must be zero at steady state if the turbulent flow 
is barotropic. 
For a barotropic fluid, the pressure is a function of the density only, 
i.e., $p=p(\rho) =p(s)$, and to prove $\langle p \nabla \cdot {\bs u} \rangle=0$, we define 
an auxiliary function, $h(s) \equiv  \int^s p(s') \exp(s-s') ds'$.
It follows from \eq{s} that ${\partial_t h}  +  \bs{u}  \cdot  \nabla  h  + h'  \nabla \cdot \bs{u}=0$, 
and, since $h(s)$ satisfies $h'(s)-h(s)=p(s)$, we have,   
\begin{equation}
\frac {\partial h}{\partial t}  +  \nabla  \cdot ( h \bs{u}) = - p \nabla \cdot \bs{u}.
\label{eq:pdV}
\end{equation}
Averaging the equation and assuming statistical stationarity and homogeneity, we 
find $\langle p  \nabla \cdot {\bs u} \rangle =0$, as claimed. This proves that 
the energy exchange by $\pdv$ work is symmetric in barotropic turbulent flows. 

For an adiabatic flow, $p \propto \rho^{\gamma}$, with $\gamma$ the ratio of specific 
heats, and $h ={p}/({\gamma-1})$ is the thermal energy. Thus, the right hand side of \eq{pdV} may be viewed as a source term 
for thermal energy, and $\langle p \cdot \nabla {\bs u} \rangle =0$  corresponds to the constancy of the average thermal energy, $\langle h \rangle$, at steady state.
The isothermal equation of state, commonly adopted in simulations for supersonic turbulence in molecular clouds, 
is another example of barotropy,  so it is expected that the average $\pdv$ work in isothermal flows 
is also zero at steady state.  
For isothermal gas, $p = \rho C_{\rm s}^2$ with $C_{\rm s}$ the constant sound 
speed, and we have $h=C_{\rm s}^2 \rho s $.  Galtier and Banerjee (2011) suggested 
that $C_{\rm s}^2 \rho s$ may be viewed as the effective thermal energy of isothermal 
gas. Clearly, $ C_{\rm s}^2 \rho s$ is not the real thermal energy of 
interstellar gas, which is given by the usual formula ${ \rho C_{\rm s}^2}/{(\gamma-1)}$. 
Defining $h=C_{\rm s}^2 \rho s $ as the effective thermal energy 
in isothermal flows was motivated by the fact that the total effective energy, defined as 
$\frac{1}{2} \rho u^2 + C_{\rm s}^2 \rho s$, is conserved by the advection and pressure terms. As an auxiliary variable, 
$h=C_{\rm s}^2 \rho s$ is helpful for understanding the energy budget in supersonic turbulence with assumed 
isothermal equation of state.

From another perspective, \eq{pdV} is also useful to understand density fluctuations in isothermal, supersonic 
turbulence. Since the sound speed is constant, $\langle h \rangle \propto \langle \rho s\rangle$, which increases with the 
width of the probability distribution function (PDF) of $s$. For example, if the PDF of $s$ 
is Gaussian with a standard deviation of $\sigma_{\rm s}$, we have $\langle \rho s\rangle = \sigma_{\rm s}^2/2$, suggesting that 
$\langle h \rangle$ may be used as a measure for the density PDF width.  
Therefore, the right hand side of \eq{pdV} can be viewed as 
a source term for density fluctuations. When the density fluctuations develop from 
the initial condition, the source term must be positive on average to widen the density PDF. 
This implies that, as the density fluctuations develop, the $\pdv$ work converts kinetic energy 
to the effective thermal energy.  But once the flow reaches steady state, the conversion stops and the average $\pdv$ 
work vanishes. 
 
The method used above in the proof for zero average $\pdv$ work  at steady state is actually not restricted to the 
specific issue of $\pdv$ work.  In general, one can prove $\langle F \nabla \cdot {\bs u} \rangle =0$ for any 
function, $F(\rho)$, of density. For a power law function, $F \propto \rho^n$, we have $\langle \rho^n \nabla \cdot {\bs u} \rangle =0$ 
for any $n$, which corresponds to the fact that the $n$-th moment, $\langle \rho^n \rangle$, of 
the density PDF is constant when the flow reaches steady state.

\subsubsection{Conditional average of the $\pdv$ work}

The result that the energy exchange by $\pdv$ work in barotropic turbulent flows is 
symmetric at steady state can be further generalized using the exact relation, \eq{exact-relation1}, in \S 2.2. 
Instead of considering the global average of the $\pdv$ 
work, we now look at the conditional average, $\langle p \nabla \cdot {\bs u} |s\rangle$,  of the $\pdv$ work 
over flow regions at each given density level, $s$. 
Clearly, if the pressure is a function of density only, we have $\langle p (\rho) \nabla \cdot {\bs u} |s\rangle =  
\langle p (\exp(s)) \nabla \cdot {\bs u} |s\rangle =  p (\exp(s)) \langle  \nabla \cdot {\bs u} |s\rangle$. 
As \eq{exact-relation1} predicted that $\langle \nabla \cdot {\bs u} |s\rangle =0$ at steady state,  it immediately follows that 
$\langle p \nabla \cdot {\bs u} |s\rangle =0$ at all values of $s$. 
We thus have reached a stronger conclusion:  At steady state, the $\pdv$ 
work does not cause a net energy exchange between kinetic and thermal 
energy in flow regions of any given density. In other words,  the $\pdv$ work is symmetric at each density 
level. 

Due to the complicated heating and cooling mechanisms, the interstellar gas is not 
really isothermal or barotropic, so the predicted symmetry of the $\pdv$ 
work based on the assumption of barotropy may 
not apply  in  general to interstellar turbulence. Nevertheless, the results in this section concerning the 
$\pdv$ work provide a useful test for the accuracy of spatial derivatives computed 
from simulation data, and we use them to demonstrate the necessity of including explicit 
viscosity in order to accurately measure the velocity divergence in simulated flows.  

\section{Testing Spatial Derivatives in Numerical Simulations}

We conducted two sets of simulations, one solving the Euler equations (without explicit viscosity), 
the other solving the Navier-Stokes equations (with explicit viscosity). Both sets of simulations
are carried out with the \emph{Dispatch} code \citep{Nordlund+18}, as it provides an efficient 
computing framework to test different fluid-dynamics solvers. The first set of experiments without 
an explicit viscous term adopted the HLLC (Harten-Lax-van Leer-Contact) approximate 
Riemann solver \citep{Toro+94}. They will be referred to as the Riemann runs. The second set of 
experiments adopted a simplified, 2nd order version of the 6th oorder solver in the \emph{Stagger Code} \citep{Galsgaard+96,Kritsuk+11,Baumann+12}, 
including the viscous term in the Navier-Stokes (N-S) equations (\eq{v}). We refer to these 
experiments with explicit viscosity as the N-S runs.

In the N-S runs, the viscous tensor, $\sigma_{ij}$, was set to $\sigma_{ij} = \rho \nu (\partial_j u_i + \partial_i u_j)$, 
where $\nu$ is the kinematic viscosity\footnote{The general form of the viscous tensor in a Newtonian fluid is 
$\sigma_{ij} = \mu (\partial_j u_i + \partial_i u_j -\frac{2}{3}\partial_k u_k \delta_{ij}) + \mu' \partial_k u_k \delta_{ij}$, 
where $\mu$ and $\mu'$ are the (dynamic) shear and bulk viscosities, respectively. For simplicity, we assumed $\mu' =\frac{2}{3}{\mu}$, 
such that $\sigma_{ij} = \mu (\partial_j u_i + \partial_i u_j)$. For interstellar conditions, $\mu'$ is typically much smaller 
than $\mu$. However, numerical stability requirements do not allow the adoption of a value 
of $\mu$ significantly smaller than $\mu'$. In this work, we focus on demonstrating the importance of including an explicit viscous 
term in order to obtain accurate spatial derivatives. Experimentation with different values for the ratio of the shear and bulk viscosities 
and exploration of the effects of that ratio on small-scale turbulent structures is left for future work.}.  
We assumed a kinematic viscosity, $\nu = \Delta x \,\mathcal{M}$, scaling linearly with cell size and Mach number,
which was sufficient to maintain numerical stability in all supersonic flows. The Reynolds number thus
increases linearly with the numerical resolution.    

For all simulations, an isothermal equation of state was adopted, and the hydrodynamic equations were evolved in a three-dimensional 
(3D) simulation box with periodic boundary conditions. The simulated flows were driven and maintained by a random, 
solenoidal force generated in Fourier space using wave numbers in the range $1<kL/2\pi<2$, where $L$ 
is the box size. Each set of simulations consists of 4 runs carried out at two numerical resolutions, $512^3$ and $1024^3$. At each resolution, 
we simulated two turbulent flows with rms Mach numbers $\mathcal{M} \simeq 3.7$ and $7.1$.  
All simulations are integrated for two sound crossing times, i.e., $2L/C_{\rm s}$, 
where $C_{\rm s}$ is  the sound speed. If the dynamical time is defined as $L/2U$, where $U$ is the 3D rms 
velocity of the flow, the simulations are integrated for $2\mathcal{M}$ dynamical times; i.e. 7.4 and 14.2 dynamical times for 
$\mathcal{M}=3.7$ and 7.1, respectively. We saved 100 snapshots per simulation, equally-spaced in time, but only used the last 
81 snapshots in the analysis, to avoid initial transients as the flow evolves from the initial conditions (uniform density and zero velocity),
and focus on steady-state statistics.

\subsection{The continuity equation}

\begin{figure}
\includegraphics[width=1\columnwidth]{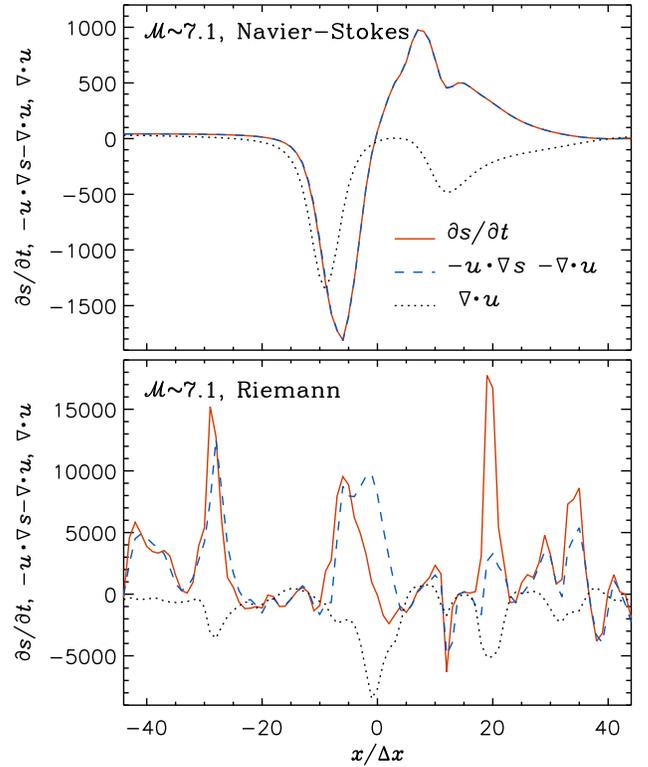}
\caption{Test of the continuity equation on a line segment 
of the  simulation grid at an arbitrarily selected snapshot. 
The chosen line segment is centered around the 
maximum density (at $x=0$) in the selected snapshot. The distance to 
the center is in units of the size, $\Delta x$, of the computational cells. 
The solid, dashed and dotted lines plot $\partial_t s$,   $- {\bs u} \cdot \nabla s - \nabla \cdot {\bs u}$, 
and $\nabla \cdot {\bs u}$, respectively, on the line segment. 
In the N-S run (top panel), the solid and dashed lines coincide, 
while the disagreement between solid and dashed lines in the Riemann run 
(bottom panel) indicates the violation of the continuity equation. 
The snapshots are taken from $1024^3$ runs with Mach 
number $\mathcal{M} =7.1$.
}
\label{figpdf}
\end{figure}

The main goal of the current work is to examine the accuracy of spatial derivatives in numerical 
simulations of supersonic turbulence. We start with a test against the continuity equation. Figure 1 
shows how well the continuity equation is satisfied in 1024$^3$ simulations of supersonic 
turbulence at Mach 7.1. The solid, dashed and dotted lines plot  $\partial_t s$,  $- {\bs u} \cdot \nabla s - \nabla \cdot {\bs u}$ and $\nabla \cdot {\bs u}$ 
on a line segment of the simulation grid at an arbitrarily selected snapshot.
The line segment is centered around the maximum density in the snapshot. 
The top and bottom panels correspond to the N-S and Riemann runs, respectively.  In the top panel, the solid and dashed lines  
coincide with each other, demonstrating that the \eq{s} is well satisfied in the N-S run.

On the other hand, significant discrepancy is seen between the solid and dashed lines 
in the bottom panel for the Riemann run. The discrepancy appears to correlate 
with the dips in the dotted line for the flow divergence, $\nabla \cdot {\bs u}$, 
which correspond to shocks.  The disagreement between the solid and dashes lines indicates that the continuity equation 
is not satisfied, and the problem is most severe around shocks. 
As pointed out in the Introduction, the spatial derivatives 
computed from simulations without explicit viscosity are controlled by the numerical algorithm adopted in the code, 
and may suffer from the artifacts induced by numerical schemes to stabilize the shocks. This is confirmed by 
the observation that \eq{s} is strongly violated around shocks. 

A comparison of the top and bottom panels of Figure 1 reveals that
structures in the N-S run are much smoother than in the Riemann run. The effective resolution or effective Reynolds number in the 
Riemann runs without explicit viscosity is significantly higher than in the simulations that evolve the 
N-S equations. This is precisely the motivation for the development of numerical codes based on 
the Euler equations, which lead to a more extended inertial range than in the simulations including 
explicit viscosity (Sytine et al.\ 2000). However, the achievement of significantly higher effective 
resolution comes with the price of losing accuracy in the
spatial derivatives computed from the simulation data.  
As a consequence, the continuity equation appears to be violated even though mass conservation is strictly enforced in the code.

\begin{figure*}
\includegraphics[width=2\columnwidth]{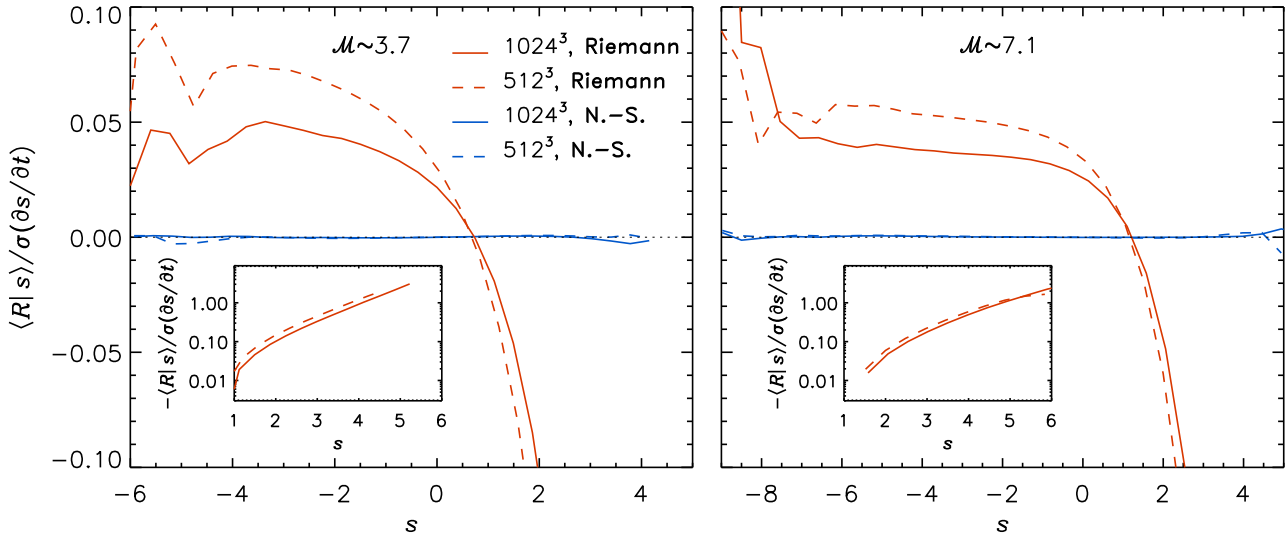}
\caption{Conditional average, $\langle R|s\rangle$, of the artificial residual of the continuity equation, normalized to the 
rms of $\partial_t s$ in all the simulated flows. Left and right panels show $\mathcal{M}=3.7$ and $7.1$, respectively. Red and 
blue correspond to the Riemann and N-S runs, 
whereas dashed and solid lines correspond to resolutions of $512^3$ and $1024^3$, respectively.  In the Riemann runs, $\langle R|s\rangle$ deviates significantly 
from zero (black dotted lines). The insets show the normalized residual at  large $s$ in the Riemann runs.}
\label{figpdf2}
\end{figure*}

\subsection{The conditional mean residual $\langle R|s\rangle$}

In Figure 2, we show the conditional average, $\langle R|s\rangle$, of the numerical residual of the 
continuity equation. As discussed in \S 2, $\langle R|s\rangle$ measures the degree to which the continuity equation is 
violated in  regions at a given density level.  All the lines in Figure 2 normalize $\langle R|s\rangle$ to $\langle (\partial_t s)^2\rangle^{1/2}$,
the rms of the rate of change of $s$ in the simulated flows. In both panels, 
the blue lines, corresponding to the N-S runs, are almost exactly zero at all $s$, demonstrating that the \eq{s} is nearly perfectly satisfied for both 
Mach numbers and both numerical resolutions. This is consistent 
with the top panel of Figure 1, which shows $R\approx 0$ in the N-S runs. 
 
In contrast, the red lines, corresponding to the Riemann runs, 
show that $\langle R|s\rangle$ deviates significantly from zero at both small and large densities.  
At intermediate to low  densities, the deviation is typically within $\simeq 10$ 
percent for both Mach numbers. Apparently, at $s\lsim 0$, $\langle R|s\rangle$ 
gets closer to 0 as the resolution increases, suggesting that at sufficiently high resolution 
the continuity equation may be better satisfied in low-density regions.

In the dense regions, the departure of $\langle R|s\rangle$ from zero 
becomes progressively stronger with increasing $s$. The problem is very severe at the 
largest values of $s$.  The insets plot $\langle R|s\rangle$  at $s\gsim 1$ with a logarithmic 
scale, showing that the departure from zero increases almost exponentially with $s$. 
At the largest density, the normalized residual reaches a value as 
large as $\simeq 1$, which means that the continuity equation is completely 
violated. As discussed earlier,  whenever numerical techniques are needed 
to stabilize shocks, numerical errors arise, and, due to the higher probability to encounter 
shocks, the denser regions are expected to suffer stronger numerical artifacts. 
This explains why the worst situation occurs at the largest $s$. 
Furthermore, at large $s$, the dependence of the normalized residual on the numerical resolution appears to 
be quite weak, especially for the $\mathcal{M}=7.1$ case (see the inset of the right panel). 
Therefore, the problem at the large densities may not be easily remedied by increasing the numerical resolution.  

The departure of $\langle R|s \rangle$  from 0 in the Riemann runs means that 
at least one of the three terms, i.e., $\langle \partial_t s|s\rangle$,  $\langle  \bs{u} \cdot \nabla s | s \rangle$, and $\langle \nabla \cdot \bs{u}| s \rangle$, 
in \eq{scondition} is nonzero. In the following subsections we examine each of the three terms to establish their relative contribution, especially at large $s$. 

\begin{figure}
\includegraphics[width=1\columnwidth]{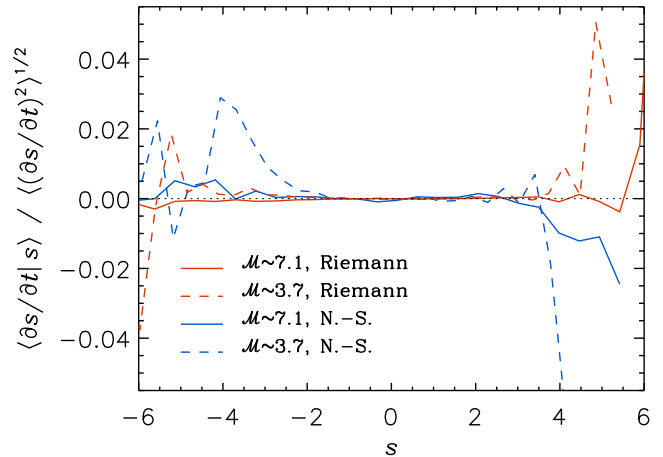}
\caption[]{Conditional mean of the time derivative of $s$, normalized to $\langle (\partial_t s)^2\rangle^{1/2}$, in $1024^3$ simulations. 
For both Riemann (red line) and N-S (blue line) runs, $\langle \partial_t s|s\rangle$ is close to zero (dotted line), with typical deviations of only a few percent. 
}
\label{figpdf3}
\end{figure}

\begin{figure*}
\includegraphics[width=2\columnwidth]{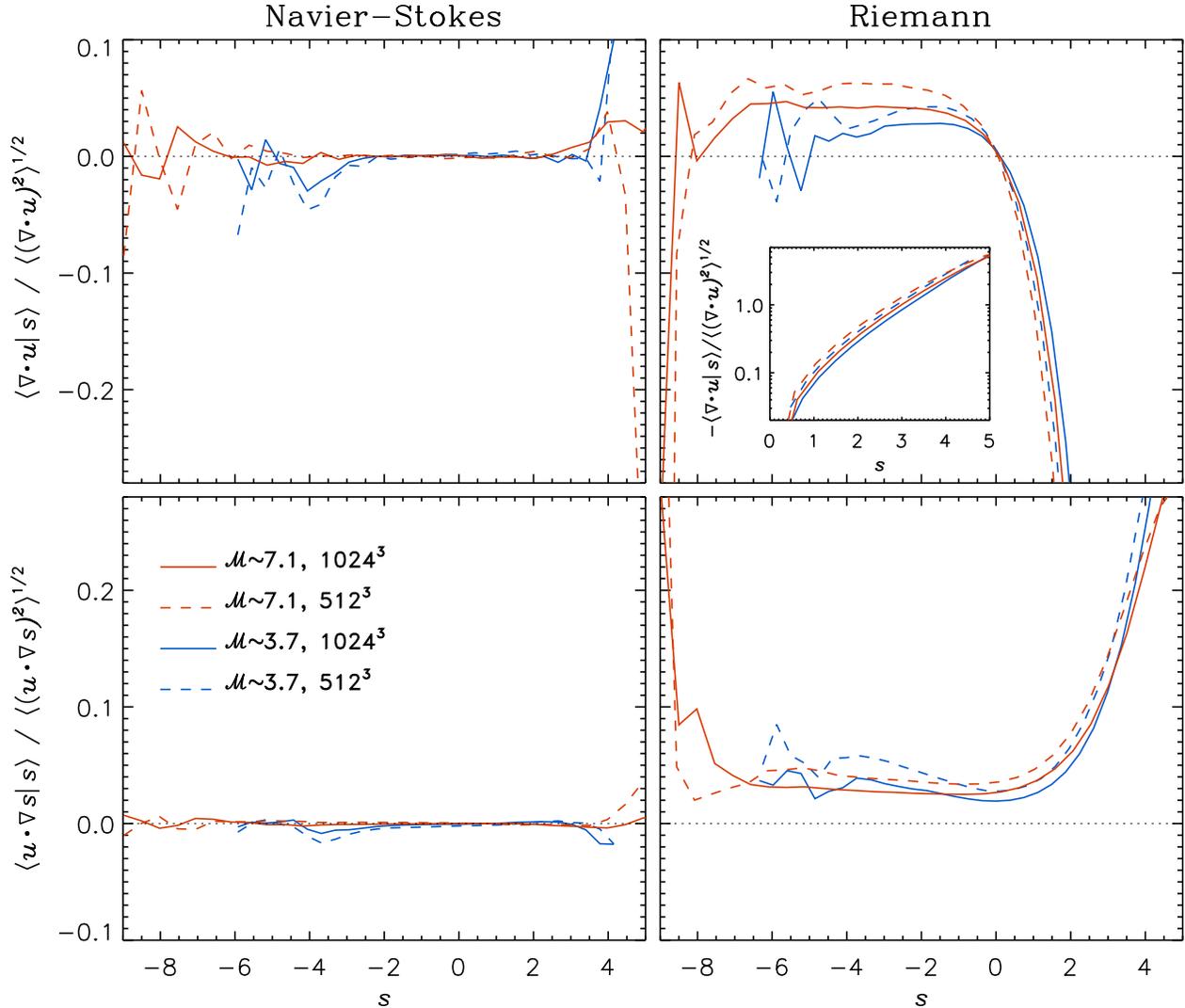}
\caption[]{Conditional mean divergence (top panels) and conditional mean advection  (bottom panels) in 
the N-S (left panels) and Riemann (right panels) runs.   Both  $\langle \nabla \cdot \bs{u}| s \rangle$  and $\langle  \bs{u} \cdot \nabla s | s \rangle$ 
are predicted to be zero (black dotted line) at steady state.  
The inset in the top right panel shows the large departure of the conditional mean divergence from zero in dense regions. 
 }
\label{figpdf4}
\end{figure*}

\subsection{The conditional mean of the time derivative of $s$}

In \S 2.1,  the conditional mean,  $\langle \partial_t s |s\rangle$,  of the time derivative of $s$  
was predicted to be exactly zero once the flow reaches 
statistically steady state.  Figure 3 plots $\langle \partial_t s |s\rangle$, normalized to the rms
of $\partial_t s$ in the $1024^3$ runs. 
It turns out that $\langle \partial_t s |s\rangle$ is close to zero in all the simulated flows, with 
typical deviations less than a few percent. The departure occurs only at extreme values of $s$ 
where the sample size is smaller, and is thus likely due to insufficient statistics.  For a given $s$ 
bin, the sample size is proportional to the probability distribution function (PDF) of $s$. 
The sample size at extreme values of $s$, corresponding to the PDF tails, increases with 
the Mach number $\mathcal{M}$ because the PDF is broader at larger $\mathcal{M}$, 
and is larger for the Riemann runs where the effective numerical resolution is higher. 
For the Riemann run at Mach 7.1, the sample size at extreme $s$ is the largest, 
and this is why the departure of the red solid line from zero is the smallest. 
On the other hand, the blue dashed line, corresponding to the N-S run at Mach 3.7, shows the 
largest noise at extreme values of $s$.  

The derivation for $\langle \partial_t s |s\rangle=0$ in \S 2.1 relies only on the assumption of steady 
state, so the finding that $\langle \partial_t s |s\rangle$ is close to zero in the simulated flows verifies 
that the snapshots used in the analysis have already reached steady state. 
Since $\langle \partial_t s |s\rangle =0$ for all flows, the departure of $\langle R|s \rangle$ from 0 found in Figure 2 in the Riemann runs must 
come from the other two terms, $\langle \nabla \cdot \bs{u}| s \rangle$  and $\langle  \bs{u} \cdot \nabla s | s \rangle$, 
in \eq{scondition}. As these two terms depend on the spatial derivatives, this implies that the spatial derivatives computed 
from the Riemann runs are inaccurate. This is consistent with the fact that the time derivative of $s$ is the value used to update
the solution for $s$, so it cannot be the source of the inaccuracy in the continuity equation (otherwise the solution itself would be inaccurate),
while the values of the spatial derivatives derived from cell centered values are not
what the HLLC solver uses to update the conserved quantities, and hence there is no \emph{a priori} reason to
expect consistency between such derivatives and the evolution of conserved quantities.
  
\subsection{Exact results of Pan et al.\ (2018)}

The conditional mean divergence, $\langle \nabla \cdot \bs{u}| s \rangle$, and conditional mean advection, $\langle  \bs{u} \cdot \nabla s | s \rangle$, 
were predicted by Pan et al.\ (2018) to be zero at steady state (\eqs{exact-relation1}{exact-relation2} in \S 2.2).  The top 
and bottom panels of Figure 3 plot  $\langle \nabla \cdot \bs{u}| s \rangle$  and $\langle  \bs{u} \cdot \nabla s | s \rangle$, 
normalized to the rms divergence, $\langle (\nabla \cdot \bs{u} )^2\rangle^{1/2}$, and the 
rms advection $\langle  (\bs{u} \cdot \nabla s )^2 \rangle^{1/2}$, respectively.  The left and right 
panels correspond to the N-S and Riemann runs, respectively.   

The prediction of Pan et al.\ (2018) is confirmed by the data from the N-S runs. In the left panels, 
we see that $\langle \nabla \cdot \bs{u}| s \rangle$ 
and $\langle  \bs{u} \cdot \nabla s | s \rangle$ are close to zero  for both Mach numbers and both resolutions.
The agreement of the N-S data with the exact result for the conditional mean advection 
is particularly impressive. As seen in  the bottom left panel, the departure of $\langle  \bs{u} \cdot \nabla s | s \rangle$ from 0 is typically 
within 1\% for all values of $s$. For the conditional mean divergence, 
the deviation from zero occurs only at extreme densities, where it oscillates around zero (see the top left panel). The oscillation 
appears to be the noise due to insufficient statistics associated with the limited sample size at extreme densities.  
Overall, the results in the left panels provide strong numerical support for the exact results derived in Pan et al.\ (2018). 

The right panels test the Riemann runs against the exact results. A similar test was conducted in Pan et al.\ (2018), who  attempted to find 
possible artifacts in simulations of supersonic turbulence due to the numerical diffusion of the density field. 
Pan et al.\ (2018) considered Mach 6 flows at different numerical resolutions, 
and their simulations were carried out using the same {\em Dispatch} code framework, in a similar 
way as the Riemann runs in the current study. The behaviors of the conditional mean divergence 
and advection shown in the right panels of Figure 4 are similar to those in Figures 1 and 2  of Pan et al.\ (2018). 

In contrast to the N-S runs, both $\langle \nabla \cdot \bs{u}| s \rangle$ and $\langle  \bs{u} \cdot \nabla s | s \rangle$ in the Riemann runs 
show strong, systematic deviations from zero, demonstrating that spatial derivatives 
based on cell centered values in these runs are 
not consistent with the actual evolution. In the top right panel, we see that $\langle \nabla \cdot \bs{u}| s \rangle$ is negative at $s\gsim 0$ 
and drops very rapidly toward large $s$. The inset shows that at high densities the amplitude of $\langle \nabla \cdot \bs{u}| s \rangle$ 
increases almost exponentially with $s$. 
A likely origin of the negative mean divergence at large densities is that when a strong convergence steepens into a shock, the Riemann solver prevents additional steepening, by effectively introducing viscosity at the cell level, while a calculation without viscosity instead suggests continued convergence of mass into those cells.
A similar argument may be applied to explain the rise of $\langle  \bs{u} \cdot \nabla s | s \rangle$ at large $s$ (see the bottom 
right panel). The advection, $\bs{u} \cdot \nabla s$, of $s$ is positive across  shocks, and sharp density jumps across shocks tend
to make $\langle  \bs{u} \cdot \nabla s | s \rangle$ based on cell centered values locally large.
Riemann solvers instead computes momentum fluxes across cell interfaces that vary more smoothly across the shock, effectively
creating results similar to viscous diffusion of momentum.

In the right bottom panel, the conditional mean advection $\langle  \bs{u} \cdot \nabla s | s \rangle$  is positive at all $s$ 
and shows a significant rise as $s$ increases above $\simeq 2$.  In Pan et al.\ (2018), the departure of $\langle  \bs{u} \cdot 
\nabla s | s \rangle$ from zero  was interpreted as related to an artificial/numerical diffusion of the density  field.   
A more correct interpretation, in light of the current result, is that it is the \emph{actual} evolution that reflects an 
effect similar to artificial/numerical diffusion, and that it is the \emph{lack} of such diffusion in direct evaluations of 
$\langle  \bs{u} \cdot \nabla s | s \rangle$ from Riemann solver solutions that produces the systematically positive values.

The interpretation of  Pan et al.\ (2018) essentially assumes that  the inaccuracy of the spatial derivatives can be 
characterized or quantified by some form of numerical diffusion, while a more precise statement is that measures of the 
inaccuracy of for example \eq{s} reflect the level of numerical diffusion that \emph{would be needed} to make the flux 
evaluations based on cell-centered values agree with the actual evolution.

Note that different Riemann solvers, with different ``sharpness" of the solutions at shocks and contact discontinuities exist.  
The HLL solver, for example, has a term that effectively diffuses both mass, momentum, and total energy across sharp interfaces. 
This leads to less sharp transitions, which would result in smaller deviations in \eq{s} than with the (better) HLLC Riemann solver. 

Similar to the results of Pan et al.\ (2018), the departure of the normalized conditional mean divergence and advection from 
zero in the Riemann runs appears to decrease with increasing resolution for both Mach 3.7 and 7.1 flows. It was speculated in Pan et al.\ (2018) 
that $\langle \nabla \cdot \bs{u}| s \rangle$ and  $\langle  \bs{u} \cdot \nabla s | s \rangle$ would approach zero in the limit of 
infinite resolution.  If so, it would be because the volume filling fraction of large discrepancies became on average smaller.

In the Mach 7.1 flows, $\langle \nabla \cdot \bs{u}| s \rangle$ and  $\langle  \bs{u} \cdot \nabla s | s \rangle$ drop and rise abruptly
towards the lowest density.  One possibility is that the abrupt behaviors  at the smallest $s$ may correspond to the numerical challenge of
handling the shocks propagating into regions of extremely low density.   
However, it is not clear whether these strong features are realistic or merely noise due to the small statistical sample at the lowest density.
The conditional means, $\langle \nabla \cdot \bs{u}| s \rangle$ and $\langle  \bs{u} \cdot \nabla s |s\rangle$, 
are related by \eq{div-adv-relation} in \S 2.2.  The equation was used in Pan et al.\ (2018) to connect the 
behaviors of $\langle \nabla \cdot \bs{u}| s \rangle$, $\langle  \bs{u} \cdot \nabla s | s \rangle$ and the PDF of $s$.  
\Eq{div-adv-relation} was derived under the assumption of statistical homogeneity only, so it holds in 
the Riemann runs, even though the spatial derivatives in those runs are inaccurate.  Based on \eq{div-adv-relation}, the abrupt drop 
of $\langle \nabla \cdot \bs{u}| s \rangle$ toward the smallest $s$ would follow from the fast rise of $\langle  \bs{u} \cdot \nabla s |s\rangle$.  

Combing Figures 2, 3 and 4, we see that in the N-S runs, \eq{s} is satisfied 
with $\langle R|s\rangle=0$ in the entire range of $s$, and all the three terms in \eq{scondition} are zero at steady state, in agreement with the predictions in \S 2.1 
and \S 2.2. In the Riemann solver runs, however, only $\langle \partial_t s |s\rangle$ is close to 
zero. Both $\langle \nabla \cdot \bs{u}| s \rangle$, $\langle  \bs{u} \cdot \nabla s | s \rangle$ show 
departures from zero, especially in the dense regions. 
Furthermore, these two terms do not add up to zero, so
$\langle R|s \rangle$ deviates significantly from zero.
In particular, $\langle R| s \rangle$ decreases very fast below zero at large $s$,  corresponding 
to the rapid drop of  $\langle \nabla \cdot \bs{u}| s \rangle$, in the same $s$ range.  
At large $s$, $\langle  \bs{u} \cdot \nabla s | s \rangle$ shows an opposite trend, but its rise 
toward large $s$ is not fast enough to compensate the rapid drop of  $\langle \nabla \cdot \bs{u}| s \rangle$.

\subsection{The $\pdv$ work at steady state}

\begin{figure}
\includegraphics[width=1\columnwidth]{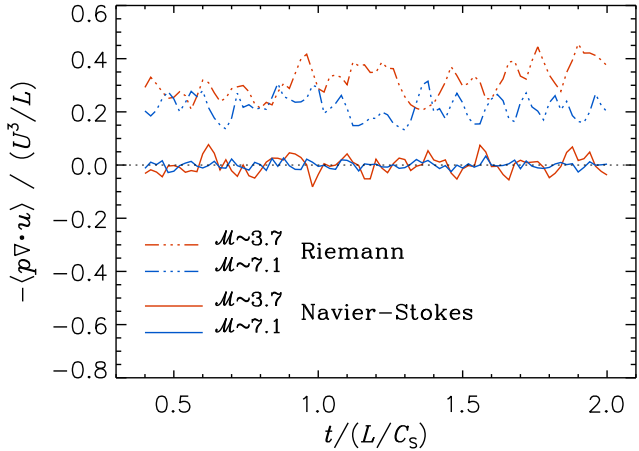}
\caption[]{Average rate of energy conversion from kinetic to thermal energy by $\pdv$ work in 1024$^3$ 
runs as a function of time. The time is in units of sound crossing time, while the conversion rate is normalized to the 
average kinetic energy dissipation rate, estimated as $U^3/L$.} 
\label{pdv1}
\end{figure}

We have shown in \S 2.3 that, in any barotropic turbulent flow, the energy transfer between 
kinetic and thermal energy by the $\pdv$ work is symmetric and the average rate of $\pdv$ work is expected to 
be zero, once the flow reaches steady state. Figure \ref{pdv1} tests the simulation data against this result. It plots the 
average rate, $-\langle p\nabla\cdot{\bs u}\rangle$, at which the $\pdv$ work converts kinetic 
energy to thermal energy as a function of time. The red and blue lines correspond to the
Mach 3.7 and 7.1 flows, and the solid and dot-dashed lines show the N-S and Riemann runs, 
respectively. The figure shows a time range from 0.4 sound crossing times 
when all the flows reach steady state to the end of the simulations at 2 sound 
crossing times. In order to check whether the $\pdv$ work plays a significant role 
in the budget of kinetic energy, we normalize it to the dissipation rate of kinetic energy, 
estimated as $U^3/L$  with  $U$ and  $L$ the 3D rms velocity and the size of the simulation 
box, respectively\footnote{The energy dissipation rate in a turbulent flow is usually estimated as 
kinetic energy, $U^2/2$, divided by the dynamical time, $\tau_{\rm dyn } = L_{\rm f}/U$, where  $L_{\rm f}$ is 
the driving length scale. Our simulated flows were driven at roughly half simulation box size, 
i.e., $L_{\rm f} \simeq  L/2$,  so the energy dissipation rate is $\simeq U^3/L$. }. 
The solid lines in Figure \ref{pdv1}  runs are very close to zero, and the amplitude of the
oscillations around zero is less than a few percent, meaning that the $\pdv$ work in 
the N-S runs does not cause a net transfer between kinetic and thermal energy, fully consistent with the 
prediction of \S 2.3.1.   

On the other hand,  $-\langle p\nabla\cdot{\bs u}\rangle$  in the Riemann runs 
appears to be positive at all times, suggesting that the energy exchange by the $\pdv$ 
work has a preferred direction. This preferential conversion was also seen in Pan \& Scannapieco (2010) 
and Kritsuk et al. (2013).  The simulations of Pan \& Scannapieco (2010) were run with the Flash code
(Fryxell et al. 2000), while the simulation used in Kritsuk et al.\ (2013) was carried out with the Enzo 
code (O'Shea et al. 2004). Both codes adopt the piecewise parabolic method (Colella \& Woodward 1984).
Like the Riemann runs in the current study, none of those previous simulations included an explicit viscous term. 
Pan \& Scannapieco (2010) computed the ratio of the energy conversion rate, $-\langle p\nabla\cdot{\bs u}\rangle$, 
by $\pdv$ work to the injection rate of kinetic energy in six simulated flows with the Mach number ranging from 1 to 6, 
and found that, at steady state, the ratio is in the range from 15\% to 30\%. 
Therefore, Pan \& Scannapieco (2010) suggested that the $\pdv$ work provides an extra channel for 
kinetic energy loss in supersonic turbulence. In the Mach 6 simulation 
used by Kritsuk et al.\ (2013), the $\pdv$ work to energy 
injection ratio is 26\%, based on the values for the average 
rate of $\pdv$ work ($-\langle p\nabla\cdot{\bs u}\rangle$=36.5) and energy 
injection rate ($\epsilon_0$ =140) provided in that study. 
The ratio of the $\pdv$ work to the energy dissipation rate in our Riemann runs is about 20-30\% (see dot-dashed lines in Figure 5), 
and considering that the energy dissipation rate is in balance with the
injection rate at steady state, this ratio is similar to those found in  Pan \& Scannapieco (2010) and Kritsuk et al.\ (2013).

\begin{figure*}
\includegraphics[width=2\columnwidth]{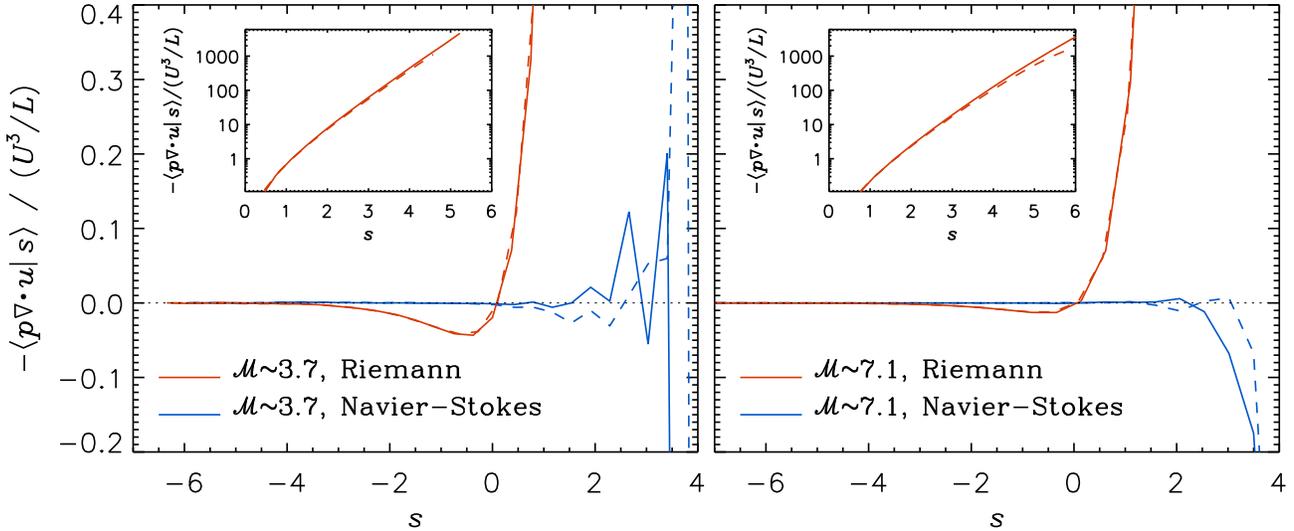}
\caption[]{Conditional mean $\pdv$ work in Mach 3.7 (left panels) and 7.1 (right panels) flows. Solid and dashed lines correspond 
to 512$^3$ and $1024^3$ runs, respectively. The insets show the sharp rise of the conditional mean $\pdv$ work in the Riemann 
runs at $s\gsim 0.5$.  In the N-S runs, departure from 0 occurs only at extremely large values of $s$ where the sample size is small. 
}
\label{pdv2}
\end{figure*}

The preferential conversion of kinetic energy to thermal energy observed in the simulations that do not include an 
explicit viscous term are clearly in contraction with our prediction in \S 2.3 that $\langle p\nabla\cdot{\bs u}\rangle =0$ at 
steady state for isothermal turbulent flows.  One possible  interpretation for 
this contradiction is that the problem is completely caused by the numerical errors in the 
divergence computed from the data. 
As shown earlier, spatial derivatives in the Riemann simulations cannot be reliably 
computed. It could be that the significant, nonzero $\pdv$ work just reflects the inaccuracy 
of spatial derivatives, and there is actually no net energy flux from kinetic energy to thermal 
energy. In other words, the apparent preferential energy conversion is just a false impression due to 
unreliable divergence computed from the data, and there is nothing wrong with the code except the spatial derivatives. 
This case is analogous to the test of the continuity equation: Although \S 3.1 \& 3.2 showed that the continuity equation is 
violated with spatial derivatives computed from the data, it does not indicate 
that the mass conservation is violated in the simulation. 

The second possibility is that  the inaccuracy in the spatial derivatives is not fully responsible for the problem, 
and in the Riemann simulations there is indeed a non-zero energy flux from kinetic energy to effective thermal 
energy by the $pdV$ work. Unfortunately, this possibility cannot be easily verified or invalidated.

Figure \ref{pdv2} shows the average rate, $-\langle p \nabla \cdot {\bs u}|s\rangle$, of the $\pdv$ work conditioned on the 
flow density. In the N-S run (the blue line), the conditional mean is found to be zero at all $s$, except at 
the largest values $s$ where the statistics is insufficient due to small sample size. 
This is in agreement with the prediction in \S 2.3.2 that, at steady state, the $\pdv$ work in barotropic 
flows is symmetric at each density level. The prediction $-\langle p \nabla \cdot {\bs u}|s\rangle =0$ simply followed from 
the fact that the pressure is a function of $s$ only in a barotropic flow and the exact result of Pan et al.\ (2018) that $\langle \nabla \cdot {\bs u}|s\rangle =0$. 

In contrast,  the red line from the Riemann solver run shows huge deviations from zero. 
For isothermal flows, the conditional mean $\pdv$ work is related to the conditional mean 
divergence by $-\langle p \nabla \cdot {\bs u}|s\rangle  = - C_{\rm s}^2 \exp(s) \langle \nabla \cdot {\bs u}|s\rangle$. 
Therefore, the behavior of  $-\langle p \nabla \cdot {\bs u}|s\rangle$ follows from that of $\langle \nabla \cdot {\bs u}|s\rangle$ shown  
in the top right panel of Figure 4. The sharp rise of $-\langle p \nabla \cdot {\bs u}|s\rangle$ above $s \simeq 0$ 
in the Riemann runs corresponds to the exponential factor and the fast drop of $\langle \nabla \cdot {\bs u}|s\rangle$ 
at large $s$ (see  Figure 4). It is this abrupt rise of $-\langle p \nabla \cdot {\bs u}|s\rangle$ 
toward large $s$ that gives the main contribution to the significant, positive overall 
average $\pdv$ work, $-\langle p \nabla \cdot {\bs u}\rangle$, as observed in Figure \ref{pdv1}. 
The huge, positive conditional mean $\pdv$ work at large $s$ in the Riemann runs further indicates 
that the spatial derivatives in the dense regions are highly inaccurate.   

In summary,  we have found that, consistent with our exact result, the $\pdv$ work in the N-S runs is symmetric at steady state, 
causing no energy flux between kinetic and thermal energy. 
If the Riemann solver solutions are analyzed using cell centered derivatives, a significant non-zero average $\pdv$ work is observed, 
which leaves the (incorrect) impression that the $\pdv$ work in supersonic turbulence preferentially converts kinetic energy to thermal energy.  

\section{Discussion}

We have found that spatial derivatives based on cell-centered values from simulations that evolve the Euler equations cannot 
be trusted. Previous works based on such simulations that involve spatial derivatives in supersonic 
turbulence need to be revisited or reinterpreted. For example, the conclusions drawn in previous studies 
from Euler simulations  concerning the $\pdv$ work in supersonic turbulence are found to be invalid.  
Kritsuk et al.\ (2013) attempted to verify the exact result of Galtier \& Banerjee (2011) 
concerning the energy cascade in the inertial range of supersonic turbulence. 
The exact result is analogous to Kolmogorov's 4/5 law for incompressible turbulence. But unlike 
Kolmogorov's 4/5 law, which only involves the 3rd order velocity structure function, the result of Galtier \& Banerjee (2011) 
includes two terms,  a ``flux" term that reduces to the 3rd order structure function in the case of incompressible flows, 
and a ``source" term that depends on the divergence. Since the divergence computed from simulations without explicit viscosity 
is unreliable, based on the discussions on the $\pdv$ work in \S 3.5, the evaluation of the source 
term in Kritsuk et al.\ (2013) using simulations with a PPM code may suffer from numerical artifact. Thus, 
the numerical analysis of the exact result of Galtier \& Banerjee (2011), in particular the analysis of the source term, 
should be re-examined using simulations that include explicit viscosity.  

In order to understand the kinetic energy cascade in compressive turbulence, 
Aluie et al.\ (2012) examined the effects of pressure dilatation using numerical simulations 
that evolved the Euler equations with a central finite-volume scheme. 
From the simulation data, they computed the pressure-dilatation cospectrum, $E^{\rm PD} (k)$, 
defined as   $E^{\rm PD} (k) = - \sum\limits_{k-0.5 \le \bs{k} <k+0.5} p^*({\bs k}) D({\bs k})$, where $p^*({\bs k})$ is the 
complex conjugate of the Fourier transform of the pressure $p$ and 
$D({\bs k})$ is the Fourier transform of the velocity divergence, $\nabla \cdot {\bs u}$. 
The integral of $E^{\rm PD} (k)$ over the $k$ space is equal to $-\langle p \nabla \cdot {\bs u}\rangle$. Aluie et al.\ (2012) stated that, as 
$K \to \infty$, the integral $\int_0^K E^{\rm PD} (k) dk$ converges to a constant, denoted as $\theta$. Clearly, $\theta$ 
corresponds to $-\langle p \nabla \cdot {\bs u}\rangle$, so 
it must be zero at steady state in their forced turbulence runs (Runs I \& III) with an isothermal equation of state. 
However, the top panels of their Figure 3 for Runs I \& III suggest that $\theta >0$, meaning that the $\pdv$ work in their simulations also appears 
to convert kinetic energy to thermal energy. In particular, in their Run III for Mach 1.25, 
the average rate of the $\pdv$ work is about 20\% of the flux of energy cascade in the inertial range. 
This is in contradiction to our result in \S 2.3 that $\theta = -\langle p \nabla \cdot {\bs u}\rangle =0$ at steady 
state. The likely reason is again that the simulations of Aluie et al.\ (2012) did not include 
viscosity, and thus the computation for the divergence is unreliable.  

\begin{figure}
\includegraphics[width=1\columnwidth]{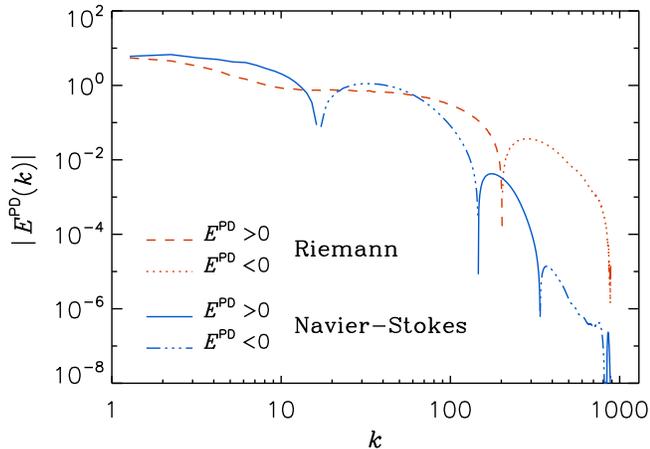}
\caption[]{Pressure-dilatation cospectrum, $E^{\rm PD} (k)$, in the 1024$^3$ simulations for the Mach 7.1 flow. The red and blue 
lines correspond to the Riemann and N-S runs. In the N-S run, the cospectrum oscillates around zero, and the integration of the cospectrum is zero, 
with the negative part canceling out the positive part.  On the other hand, the cospectrum in the Riemann run does not integrate to zero.  
 }
\label{cospectrum}
\end{figure}

The problem can also be seen from the top panel of their Figure 2 for the cospectrum, $E^{\rm PD} (k)$. Apparently, 
$E^{\rm PD} (k)$ in their forced runs does not integrate to zero. 
The cospectrum computed from our simulation data shows that $E^{\rm PD} (k)$ in the N-S runs is qualitatively different from that 
in the Riemann runs. As seen in our Figure \ref{cospectrum}, in the Riemann runs, $E^{\rm PD} (k)$ is positive in most $k$ range, 
and becomes negative only towards the dissipation range. The cospectrum in the Mach 1.45 run of  Aluie et al.\ (2012) shows a similar behavior. 
In contrast, the cospectrum in the N-S runs oscillates around zero, and its integral is zero as the negative part of the spectrum cancels 
out the positive part. This suggests that the cospectrum computed by Aluie et al.\ (2012) suffers from inaccuracies in the divergence
evaluated from their simulation, and the effects of pressure dilatation interactions on the kinetic energy budget in compressible turbulence 
should be reexamined using simulations that include explicit viscosity. 

Large differences in $E^{\rm PD} (k)$ between the Riemann and N-S runs are found also in inertial-range wave numbers 
(Figure \ref{cospectrum}), suggesting that the inaccurate evaluation of spatial derivatives from solutions of turbulent flows computed 
without explicit viscosity may even affect inertial-range statistics. 
One might attribute this qualitative difference in the inertial range solely to the inaccurate divergence 
in the Riemann runs, and expect that no such problem arises for flow quantities that do not involve gradients.  
However, we have verified the same problem arises when computing the cospectrum in terms of the Fourier transform, 
$u_i({\bs k})$, of the velocity field as  
$E_i^{\rm PD} (k)  = -i \sum\limits_{k-0.5 \le \bs{k} <k+0.5} k_i p^*({\bs k}) u_i({\bs k})$. 
Therefore, the difference in $E^{\rm PD} (k)$ between the Riemann and N-S runs 
indicates that the pressure-velocity cospectrum, $\langle p^*({\bs k}) u_i({\bs k})\rangle$ (or equivalently 
the cross correlation function, $\langle p({\bs x})u_i ({\bs x} + {\bs r})\rangle$) in the Riemann runs also suffers 
from numerical artifacts that extend to inertial-range scales. The pressure-velocity cospectrum does not involve any 
spatial gradients, suggesting the existence of inaccuracies in the post-processing of Euler simulations
even when spatial derivatives are not involved. Future work should address the extent to which such inaccuracies
affect other inertial-range diagnostics in Euler simulations.

Finally, we point out that the study of the viscous dissipation of kinetic energy in turbulent flows requires an accurate computation of the 
velocity gradient (e.g.\ Pan \& Padoan 2009, Pan et al.\ 2009), so the statistics of the dissipation rate in supersonic turbulence should 
be re-analyzed using simulations that include explicit viscosity. This represents a significant hurdle for such study, because we generally 
aim at characterizing the viscous dissipation in the limit of very large Reynolds number, which is difficult to achieve with N-S runs. 

\section{Conclusions}

We examined the accuracy of spatial derivatives in numerical simulations of supersonic turbulence. 
We conducted two sets of simulations using the \emph{Dispatch} code framework, one based on a finite-volume 
method to solve the Euler equations and the other based on a finite-difference method to solve the 
Navier-Stokes equations. We tested them against a number of criteria based on the continuity equation,  
including some exact results derived from the continuity equation at statistically steady state.  We summarize 
our conclusions as follows:  
\begin{enumerate}
\item The spatial derivatives based on the N-S runs are accurate and satisfy all the criteria adopted in our 
study. In particular,  the data from N-S runs confirm the exact results derived by Pan et al.\ (2018), 
i.e., the conditional mean divergence, $\langle \nabla \cdot {\bs u}|s\rangle$, and the conditional mean advection 
$\langle  {\bs u} \cdot \nabla s|s\rangle$ vanish at statistically steady state.

\item
Without an explicit physical viscosity, the structure of discontinuities in the Riemann solver is controlled by numerical schemes 
involving flow corrections designed to achieve numerical stability. This induces errors in finite-difference derivative expressions, 
even if  the code enforces the conservation laws (in integral form). As a consequence, the Riemann solver runs fail all 
the tests that require accurate evaluation of spatial derivatives

\item In the Riemann solver runs, the continuity equation appears to be violated with spatial derivatives computed from the data, 
especially in high-density regions where most shocks inhabit, even though mass conservation is strictly enforced in the code. 

\item 
In the Riemann solver runs, $\langle \nabla \cdot {\bs u}|s\rangle$ and $\langle  {\bs u} \cdot \nabla s|s\rangle$ 
deviate significantly from zero at large densities.  These deviations further illustrate the inaccuracy of using 
cell-centered finite-difference spatial derivatives when post-processing data from simulations that do not include explicit viscosity.  

\item
We have shown that the energy exchange between kinetic and thermal energy by $\pdv$ work is symmetric in isothermal, 
supersonic turbulence once the flow reaches steady  state. This analytical result is confirmed by the N-S runs. 

\item
The inaccuracy of spatial derivatives based on the Riemann solver runs gives the 
incorrect impression that the $\pdv$ work tends to preferentially convert kinetic energy to thermal 
energy. This problem also exists in the interpretation of previous simulations using other numerical codes, 
indicating that the unreliability of cell-centered spatial derivatives is a general issue for all simulations 
that do not include explicit viscosity.   
Furthermore, we  found that the pressure-dilatation cospectrum in the N-S and Riemann runs shows large qualitative 
differences also at inertial-range scales.

\end{enumerate}

Our work suggests that studies involving spatial gradients in supersonic turbulence must be carried out and interpreted with caution. 
In principle, one may need to include explicit viscosity in the simulation in order to obtain accurate spatial derivatives in post-processing analyses. 
The extent to which diagnostics of inertial-range dynamics are also affected, as in the example of the pressure-dilation cospectrum reported here, will be further 
investigated in future studies.

\acknowledgements
LP acknowledges support from the Youth Program of the Thousand Talents Plan in China.
PP acknowledges support  by the Spanish MINECO under project AYA2017-88754-P. 
The work of {\AA}N was supported by grant 1323-00199B from the Danish Council for Independent 
Research (DFF), and by the Centre for Star and Planet Formation, which is funded by the Danish National Research Foundation (DNRF97).
Storage and computing resources at the University of Copenhagen HPC centre, funded in part by Villum Fonden (VKR023406), were used 
to carry out the simulations. We thankfully acknowledge the computer resources at MareNostrum and the technical support provided by 
Barcelona Supercomputing Center (AECT-2018-3-0019).

\end{document}